\documentclass[%
reprint,
 amsmath,amssymb,
 aps,
pra,
]{revtex4-2}

\usepackage[T1]{fontenc}
\usepackage[utf8]{inputenc}
\usepackage{mathptmx}
\usepackage{physics}

\usepackage{graphicx}
\usepackage{dcolumn}
\usepackage{bm} 
\usepackage[colorlinks=True,urlcolor=blue,citecolor=blue,linkcolor=blue]{hyperref}

\usepackage{amsthm}
\theoremstyle{definition}

\theoremstyle{theorem}

\usepackage{tikz}
\usetikzlibrary{angles,quotes}
\usetikzlibrary{quantikz2}
\usepackage{soul}

\def\red{\textcolor{black}}

\def\angle{\frac{\theta}{2}}

\usepackage{float}
\usepackage{booktabs}
\usepackage{microtype}

\usepackage{ragged2e}

\begin{document}
\raggedbottom

\preprint{APS/123-QED}

\title{Quasi Inverse of Qubit Channels for Mixed  States}
\author{Muhammad Faizan}
\author{Muhammad Faryad}
 \email{muhammad.faryad@lums.edu.pk}
 \affiliation{Department of Physics, Lahore University of Management Sciences, Lahore 54792, Pakistan.}

%


\date{\today}

\begin{abstract}
 We found the quasi inverse of qubit channels as a unitary map, $\mathcal{E}^i$, by minimizing the average trace distance between the input state to the channel and the output of the quasi inverse channel for arbitrary qubit channel $\mathcal{E}$ and for arbitrary input states. The channel $\mathcal{E}$ was assumed completely positive and trace-preserving. To find the quasi inverse for mixed states, we proposed an alternative definition of the quasi inverse based on the mean square of the trace distance (MSTD) of a channel. The definition based on the trace distance allowed easy generalization of the quasi inverse to mixed input states. The quasi inverse of the Pauli,  generalized amplitude damping, mixed unitary, and tetrahedron channels calculated based on the MSTD agreed with the one computed using average fidelity in the special case of input states being pure. 
\end{abstract}

\maketitle
\section{Introduction}
Qubits are the building blocks of quantum computers, but they are quite vulnerable to unwanted interactions with the environment. This interaction leads to the decoherence   of the quantum state and makes the computations erroneous on quantum computers. This interaction process can be modeled as a quantum channel \cite{nielsen2010}. In general, quantum channels are irreversible operations \cite{karimipour2020qubit,shahbiegi2021finite,karimipour2022on}. However, the idea of quasi inversion of the quantum channel was recently introduced \cite{karimipour2020qubit}  to reverse a part of the effect of that channel. The concept of quasi inverse is essential for enhancing the fidelity of quantum gates and reducing errors in quantum computations. This idea is also helpful in quantum inverse problems, such as state estimation, where quasi inverses are used to improve efficiency and resilience \cite{cao2022neural}. 

The quasi inverse  is  defined so as to  increase  the average fidelity of the channel in Ref. \cite{karimipour2020qubit}. The derivation of the quasi inverse was based on maximizing the average fidelity between the input state of the channel and the output state of the quasi inverse of that channel. However, the derivation was restricted to the case when the input state was a pure state.
In the present work, we propose an alternative way of defining the quasi inverse using the trace distance that allows us to consider mixed input state.

 The distance measures between the quantum states play a pivotal role in quantum information science. This is used to compute quantum correlations, entanglement, and coherence as discussed in \cite{vesperini2023ent}. 
The distance measures come into play for several critical tasks in quantum information processing. They help us gauge and distinguish different states of entanglement in bipartite and multipartite systems, especially when the observable set is not closed under products \cite{ghosh2018quantum}.  Among these measures, the trace distance and fidelity are frequently used \cite{nielsen2010,wilde2013}. However, fidelity is usually easier to calculate only when one of the state is a pure state. On the other hand, the trace distance does not suffer from this drawback. Therefore, to extend the concept of quasi inverse for mixed input states, we propose the mean square trace distance (MSTD): {the square of the trace distance averaged over all possible input states.}

The structure of the paper is as follows: {In Section \ref{sec-mstd}, we construct the MSTD for a channel from the definition of the trace distance between two density matrices.} In Section \ref{sec2}, we define the quasi inverse {using the MSTD} and the general derivation of quasi inverse for qubit channels is presented in Section \ref{sec3}. In Section \ref{sec4}, we derive the quasi inverses of a few  qubit channels  and concluding remarks are presented in Section \ref{sec5}.

\section{Mean square trace distance (MSTD)}\label{sec-mstd}

{Consider two single-qubit mixed states $\rho$ and $\zeta$. The  density matrices $\rho$ and $\zeta$ can be represented as geometric vectors in the unit Bloch ball with their position vectors $\mathbf r$ and $\mathbf z$ having components $r_i = {\rm Tr}(\rho\sigma_i)$ and $z_i = {\rm Tr}(\zeta\sigma_i)$, $i=1,2,3,$ where $\sigma_i$ are the Pauli operators, $0\leq r\leq 1$, and $0\leq z\leq 1$. Therefore, we can express the density matrices in terms of the Bloch vectors as $$\rho =\frac{1}{2} (I + \mathbf{r}\cdot \mathbf{\sigma})$$ and $$\zeta =\frac{1}{2}  (I + \mathbf{z}\cdot \mathbf{\sigma})$$ with $  \mathbf{\sigma}$ being the vector of the Pauli operators and $I$ being the identity operator.}

{The trace distance between two single-qubit mixed states $\rho$ and $\zeta$ is defined as \cite{nielsen2010}
\begin{eqnarray}
	D(\rho, \zeta) &=&\frac{1}{2}{\rm Tr}\,\abs{\rho-\zeta}\nonumber \\
	&=&\frac{1}{4}{\rm Tr}\abs{(\mathbf{r} - \mathbf{z})\cdot\mathbf{\sigma}}\nonumber \\
	&=&\frac{1}{2}\abs{\mathbf r-\mathbf{z}}\label{td}
\end{eqnarray}
where we used the fact that the eigenvalues of $(\mathbf{r} - \mathbf{z})\cdot\sigma$ are $\pm\abs{\mathbf{r} - \mathbf{z}}$.}

Now consider a qubit channel $\mathcal{E}$ as shown schematically in Fig. \ref{schematic1}. The MSTD of the channel is defined as
\begin{equation}\label{mstdo}
	\overline{D^2}(\mathcal{E}) =\int D^2\left(\rho,\mathcal{E}(\rho)\right)\,\mathrm{d}\rho\,,
\end{equation}
where the integration is to be performed over all single-qubit mixed states.

\begin{figure}[!h]
	\centering
\begin{tikzpicture}
\node[scale=1.5]{
\begin{quantikz}
		\lstick{$\rho$} &
  \gate{\mathcal{E}}  & \rstick{$\mathcal{E}(\rho)$} 
\end{quantikz}   
};
 \end{tikzpicture}
	\caption{The schematic showing the qubit channel $\mathcal{E}$ with mixed input state $\rho$ and output state $\mathcal{E}(\rho)$.} \label{schematic1}
\end{figure}

The quantum channel $\mathcal{E}$ can be represented as an affine transformation of the Bloch vectors of the input density matrices to the Bloch vectors of the output density matrices as \cite{nielsen2010}
\begin{equation}
    \mathbf{r} \longrightarrow \mathbf{z}= M\mathbf{r} + \mathbf{c}\,,
\end{equation} 
where ${M}$ is a real $3\times 3$ matrix and $\mathbf{c}$ is a real vector in $\mathbb{R}^3$ given by
\begin{eqnarray}
M_{ij} &=& \frac{1}{2}\text{Tr}(\sigma_i \mathcal{E}(\sigma_j))\,, \quad i,j=1,2,3,\\
c_i &=& \frac{1}{2}\text{Tr}(\sigma_i \mathcal{E}(I))\,, \quad i=1,2,3\,.
\end{eqnarray}
Hence, a quantum channel can be characterized by the pair $({M},\mathbf{c})$. Therefore, using Eq. (\ref{td}), the definition (\ref{mstdo}) can be recast as
\begin{eqnarray}\label{mstd}
	\overline{D^2}(\mathcal{E}) &=&\frac{1}{4}\int\abs{\mathbf{r}-\mathbf{z}}^2~\mathrm{d}^3\mathbf{r}\,,\\
	&=&\frac{1}{4}\int\abs{\mathbf{r}-\left(M\mathbf{r} + \mathbf{c}\right)}^2~\mathrm{d}^3\mathbf{r}\,,
\end{eqnarray}
where the integral is taken over the unit Bloch ball with $r\leq1$ to average over all mixed input states $\rho$. Let us note that this integral will be over the surface of the Bloch sphere if one wishes to consider only the pure input states $\rho=\ket{\psi}\bra{\psi}$. However, in this paper we assume that the input states can be mixed as well as pure.

\section{Quasi inverse in terms of MSTD}\label{sec2}
Consider a general single-qubit channel $\mathcal{E}$ that converts a mixed state $\rho$ to $\mathcal{E}(\rho)$. Suppose further  {that another channel $\mathcal{E}^\prime$ transforms $\mathcal{E}(\rho)$ to $\rho^\prime= (I + \mathbf{r}^\prime\cdot \mathbf{\sigma})/2=\mathcal{E}^\prime\circ\mathcal{E}(\rho)$ such that
\begin{equation}\label{eq2p}
        \overline{D^2}\big(\mathcal{E}^\prime\circ\mathcal{E}\big) \leq \overline{D^2}\big(\mathcal{E}\big) \,.
	\end{equation}
 as shown schematically in Fig. \ref{schematic2}. 
 The quasi inverse $\mathcal{E}^i$ is a channel such that
 \begin{equation}
 \mathcal{E}^i = \max\limits_{\Delta\overline{D^2} }\,\,\mathcal{E}^\prime 
 \end{equation}
 	where 
\begin{equation}
\Delta\overline{D^2} = \overline{D^2}\big(\mathcal{E}\big)-\overline{D^2}\big(\mathcal{E}^\prime\circ\mathcal{E}\big) \,.\label{max}
\end{equation}}

\begin{figure}[!h]
	\centering
\begin{tikzpicture}
\node[scale=1.5]{
\begin{quantikz}
		\lstick{$\mathcal{E}(\rho)$} &
   \gate{\mathcal{E}^\prime} & \rstick{$\mathcal{E}^\prime\circ\mathcal{E}(\rho)$} 
\end{quantikz}   
};
 \end{tikzpicture}
	\caption{The schematic showing quasi inverse of the qubit channel $\mathcal{E}$.} \label{schematic2}
\end{figure}

Therefore, the channel {$\mathcal{E}^\prime$} will be quasi inverse if it {maximizes $\Delta\overline{D^2}(\mathcal{E}^\prime\circ\mathcal{E})$} for a given channel $\mathcal{E}$. Let us note that it is  equivalent to either minimize \( \overline{D^2}(\mathcal{E}^\prime\circ \mathcal{E}) \) or maximize $\Delta \overline{D^2}$ over all CPTP maps \( \mathcal{E}^i \). \red{However, maximizing $\Delta \overline{D^2}$ is preferred over minimizing $\overline{D^2}(\mathcal{E}^\prime \circ \mathcal{E})$, as both yield equivalent quasi inverses, but the former produces equations with a clear and interpretable structure, consistent with Ref. \cite{karimipour2020qubit} and the calculations are simpler.} Therefore, similar to the definition of the quasi inverse based on fidelity \cite{karimipour2020qubit}, we define the quasi inverse $\mathcal{E}^i$ as a channel that reduces the MSTD between its output state and the  input state of the original channel $\mathcal{E}$  by at least the same or more than any other channel $\mathcal{E}^\prime$, i.e.,
\begin{equation}\label{eq2}
        \overline{D^2}\big(\mathcal{E}^i\circ\mathcal{E}\big) \leq \overline{D^2}\big(\mathcal{E}^\prime\circ\mathcal{E}\big) \qquad \forall \qquad \mathcal{E}^\prime \,.
\end{equation}

It was proved in Ref. \cite{karimipour2020qubit} that the inverse channel can only be a unitary operator \textit{if} the input state is assumed to be a pure state and the inverse is defined via average fidelity. Let us note that this can be assumed to be true in general as well about the quasi inverse as follows: The action of a quantum channel can be decomposed into rotation and scaling of the Bloch vector of the input state \cite{nielsen2010}. Since any trace-preserving quantum channel cannot increase the length of the Bloch vector, the optimal quantum operation to reverse the effect of rotation and scaling can only be rotation embodied by a unitary transformation.  Therefore, the inverse channel essentially rotates the state towards the input state to reduce its distance from the original state without changing the length of  the Bloch vector. So, we can assume that the optimal quasi inverse can only be a unitary operation.

\section{General Derivation}\label{sec3}
Assuming the quasi inverse as a unitary operator, we can write
\begin{equation}
    \mathcal{E}^i(\rho) = V \rho V^\dagger\,,\label{qidef}
\end{equation}
where $V$ is a unitary operator that can be written as
\begin{equation}\label{defV}
V = x_0\mathbb{I} + i\mathbf{x}\cdot \mathbf{\sigma}
\end{equation}
with $x_0$ a real number, $\mathbf{x}$ real vector, and $x_0^2 + \mathbf{x}\cdot\mathbf{x} = 1$.

Finding the quasi inverse is essentially finding four real parameters $(x_0,\mathbf{x})$ that specifies the unitary $V$. To do this, we can set up a constrained optimization problem to maximize $\Delta \overline{D^2}$ (the decrease in the MSTD) 
 over all unitary maps, i.e., maximizing over the real parameters $(x_0,\mathbf{x})$, subject to the constraint $x_0^2 + \mathbf{x}\cdot \mathbf{x}=1$, for a given channel $\mathcal{E}$.

Using Eq. \eqref{mstd}, the MSTD {of the channel $\mathcal{E}$} is found as
\begin{equation}
 \overline{D^2}(\mathcal{E}) = \frac{1}{20} \left( \text{Tr}(MM^\dagger) - 2\text{Tr}(M) + 3 \right) + \frac{1}{4}\abs{\mathbf{c}}^2\,.\label{ms1}
\end{equation}
To evaluate the MSTD integral, the following identities were used:
\begin{equation}\label{vol-integral}
\int r_i~\mathrm{d}^3\mathbf r=0\,, \qquad \int r_ir_j~\mathrm{d}^3\mathbf r=\frac{1}{5}\delta_{ij}\,,
\end{equation}
\begin{equation}
\begin{gathered}
\mathrm{d}^3\mathbf r  = r^2 \sin\theta~\mathrm{d}r~\mathrm{d}\theta~\mathrm{d}\phi\,, \\
\mathbf{r} = ( r\sin\theta\cos\phi , r\sin\theta\sin\phi , r\cos\theta )\,, \\
r\in [0,1]\,,\quad \theta\in [0,\pi]\,,\quad \phi\in[0,2\pi]\,.
\end{gathered}
\end{equation}

The MSTD of combined channel $\mathcal{E}^i\circ\mathcal{E}$ is given by
\begin{equation}
 \overline{D^2}\big(\mathcal{E}^i\circ\mathcal{E}\big) = \frac{1}{20} \left( \text{Tr}(NN^\dagger) - 2\text{Tr}(N) + 3 \right) + \frac{1}{4}\abs{\mathbf{u}}^2\,,\label{ms2}
\end{equation}
where
\red{
\begin{align}
	N & = M^i M \label{matrix_N} \\
	\mathbf u &= M^i \mathbf c + \mathbf c^i
\end{align}
since the composition of a quantum channel is reflected in the composition of the transformation matrices \cite{nielsen2010}
}
\begin{equation}
\mathcal{E}_2\circ\mathcal{E}_1\equiv  (M_2M_1 , M_2\mathbf{c}_1+\mathbf{c}_2)
\end{equation}
with $\mathcal{E}_{1,2} \equiv (M_{1,2},\mathbf{c}_{1,2})$.

{
The affine map for the quasi inverse channel (\ref{qidef}) can be written using simple algebra as
\begin{equation}\label{m_i}
	\begin{gathered}
		M^i = \begin{pmatrix}
			1 -2 (x_{2}^{2} + x_{3}^{2}) & 
			2 (x_{0} x_{3} +  x_{1} x_{2}) & 
			- 2 (x_{0} x_{2} -  x_{1} x_{3}) \\
			- 2 (x_{0} x_{3} - x_{1} x_{2}) & 
			1 - 2 (x_{1}^{2} + x_{3}^{2}) & 
			2 (x_{0} x_{1} +  x_{2} x_{3}) \\
			2 (x_{0} x_{2} +  x_{1} x_{3}) & 
			- 2 (x_{0} x_{1} - x_{2} x_{3}) &  
			1 - 2 (x_{1}^{2} + x_{2}^{2})
		\end{pmatrix} \,,\\
		\mathbf{c}^i = \begin{pmatrix}
			0 & 0 & 0
		\end{pmatrix}^T\,.
	\end{gathered}	
\end{equation}
The last equality is due to the fact that the quasi inverse is taken to be a unitary channel.	
}

Substituting Eqs. (\ref{ms1}) and (\ref{ms2}) in Eq. (\ref{max}) and maximization process gives out the quasi inverse for any qubit channel with mixed input states. 

\section{Examples}\label{sec4}
To find the quasi  inverse of  specific channels, we consider the same example channels as in Ref. \cite{karimipour2020qubit} for comparison of the results. {However, we begin with a sanity check and find the quasi inverse of a general unitary channel since its inverse can be exactly written using the adjoint of unitary operator defining the channel. This will also illustrate the process of finding quasi inverse clearly.}

\subsection{General Unitary Operator}
{
A general unitary operator is a rotation of the qubit state by some angle $\theta$ about an arbitrary axis $\hat{n}$ and can be written as
\begin{equation}
	U = e^{-i\theta \mathbf n \cdot \sigma/2} = \cos\left(\frac{\theta}{2}\right) I - i \sin\left(\frac{\theta}{2}\right) \left[n_1 \sigma_1 + n_2 \sigma_2 + n_3 \sigma_3\right]
\end{equation}
where $\mathbf n = \left( n_1,n_2,n_3 \right)$ is a real unit vector in $\mathbb R^3$.
}
{
The general unitary channel defined as
\begin{equation}
	\mathcal{E}(\rho) = U \rho U^\dagger
\end{equation}
can also be equivalently defined using the affine map like that in Eq. (\ref{m_i}) as
\begin{equation}\label{affine_general}
	M = \begin{pmatrix}
		\beta_x &
		\alpha_{xyz} &
		\alpha_{xzy} & \\
		\alpha_{xyz} & 
		\beta_y &
		\alpha_{xzy}  \\
		\alpha_{xzy} &
		\alpha_{yzx} &
		\beta_z
	\end{pmatrix}
\end{equation}
where $\mathbf{c} =(0, 0, 0)^T$, $\beta_i = 2n_i^2 \sin^2\left(\angle\right)+\cos(\theta)$ and $\alpha_{ijk} = 2n_in_j\sin^2\left(\angle\right)-n_k\sin(\theta)$. Using Eqs. \eqref{ms1}, \eqref{ms2}, \eqref{m_i}, and \eqref{affine_general} in Eq.  \eqref{max} we get
\begin{equation}
\begin{gathered}
\Delta\overline{D^2} = \frac{2}{5} \left[
\omega_x x_1^2 + \omega_y x_2^2 + \omega_z x_3^2 \right. \\
+ 2\epsilon_x x_0x_1
+ 2\epsilon_y x_0x_2
+ 2\epsilon_z x_0x_3 \\
\left.
2\mu_{xy} x_1x_2 +
2\mu_{xz} x_1x_3 +
2\mu_{yz} x_2x_3
\right]\,,
\end{gathered}
\end{equation}
where $\epsilon_i = \frac{1}{2}n_i \sin(\theta)$, $\omega_i = n_i^2 \sin^2\left(\angle\right) +\sin^2\left(\angle\right)-1$, and $\mu_{ij}=n_in_j\sin^2\left(\angle\right)$.
}
{
This allows us to write the above expression in matrix representation given by:
\begin{equation}\label{d2matrix}
	\Delta\overline{D^2} = \frac{2}{5}\begin{pmatrix}
		x_0 & \mathbf{x}^T
	\end{pmatrix} Q \begin{pmatrix}
		x_0 \\
		\mathbf{x}
	\end{pmatrix}\,,
\end{equation}
where
\begin{equation}
	Q = \begin{pmatrix}
		0 & \epsilon_x & \epsilon_y & \epsilon_z \\
		\epsilon_x & \omega_x & \mu_{xy} & \mu_{xz} \\
		\epsilon_y & \mu_{xy} & \omega_y & \mu_{yz} \\
		\epsilon_z & \mu_{xz} & \mu_{yz} & \omega_z
	\end{pmatrix}\,.
\end{equation}
The eigenvalues of $Q$ are given by:
\begin{equation}
	\begin{aligned}
		\lambda_1=\lambda_2=\lambda_3 & = -\cos^2\left(\angle\right)  ,\quad  \lambda_4  = \sin^2\left(\angle\right)
	\end{aligned}
\end{equation}
Thus, the decrease in the MSTD is given by:
\begin{equation}
	\begin{aligned}
		\Delta\overline{D^2} & = \frac{2}{5}\max\left(\lambda_1,\lambda_2,\lambda_3,\lambda_4\right) \\
		&= \frac{2}{5} \sin^2\left(\angle\right)\\ 
		& = D^2(\mathcal{E})
	\end{aligned}
\end{equation}
The eigenstate corresponding to the largest eigenvalue is our solution as explained in the Appendix \ref{appA}. The largest eigenvalue $\lambda_{\max}=\lambda_4$ corresponds to the eigenstate 
\begin{equation}
	\ket{\lambda_{\max}} = \begin{pmatrix}
		\cos\left(\angle\right) \\[2mm]
		n_x \sin\left(\angle\right) \\[2mm]
		n_y \sin\left(\angle\right) \\[2mm]
		n_z \sin\left(\angle\right)
	\end{pmatrix}\,.
\end{equation}
Thus the quasi inverse is:
\begin{equation}
\begin{aligned}
V  & = \cos\left(\angle\right)I + i\sin\left(\angle\right)\left(\mathbf n \cdot \mathbf{\sigma}\right) \\
& = e^{i\theta \mathbf{n}\cdot \mathbf{\sigma}} \\
& = U^\dagger
\end{aligned}
\end{equation}
after using the fact that $(x_0,\mathbf x)=(x_0,x_1,x_2,x_3)=\ket{\lambda_{\max}} $ and Eq. (\ref{defV}).
}

\subsection{The Pauli Channel}
Pauli channel is ubiquitous in quantum information processing devices and is used as the most common model for qubit noise  in computing hardware. It can be specified using Kraus operators as 
\begin{equation}
    \mathcal{E}(\rho) = p_0\rho + \sum_{i=1}^3 p_i \sigma_i \rho \sigma_i
\end{equation}
with $p_i\geq 0$ and $\sum_{i=0}^3p_i = 1$.

{The affine map for Pauli channel can be computed as
\begin{equation}\label{pauli_M}
\begin{gathered}
	M = \begin{pmatrix}
		\alpha_{23} & 0 & 0 \\
		0 & \alpha_{13} & 0 \\ 
		0 & 0 & \alpha_{12}
	\end{pmatrix} \quad , \quad 
\mathbf{c} = \begin{pmatrix}
				0 & 0 & 0
			\end{pmatrix}^T\,,
\end{gathered}
\end{equation}
where $\alpha_{ij} = 1-2p_i - 2p_j $.}

Using Eqs. \eqref{ms1}, \eqref{ms2}, \eqref{m_i}, and \eqref{pauli_M} in Eq.  \eqref{max} we get
\begin{equation}\label{deltad2pauli}
	\begin{gathered}
		\Delta\overline{D^2} = \frac{2}{5}\Big( -  p_{1} x_{0}^{2} + 
		 p_{1} x_{1}^{2} + 
		 p_{1} - 
		 p_{2} x_{0}^{2} + 
		 p_{2} x_{2}^{2} + 
		 p_{2} - \\ 
		 p_{3} x_{0}^{2} + 
		  p_{3} x_{3}^{2} + 
		 p_{3} + \frac{3}{4} x_{0}^{2} - 
		 \frac{1}{4} x_{1}^{2} - 
		 \frac{1}{4} x_{2}^{2} - 
		 \frac{1}{4} x_{3}^{2} - \frac{3}{4} \Big)\,.
	\end{gathered}
\end{equation}

Using the identities $x_0^2 + \mathbf x\cdot \mathbf x = 1$ and $\sum_{i=0}^3p_i=1$, \red{the above expression takes the form of Eq. (\ref{d2matrix}) with}
\begin{equation}
\begin{gathered}
Q = \text{diag}\left(0 , p_1-p_0 , p_2-p_0, p_3-p_0\right)\,.\label{Qpauli}
\end{gathered}
\end{equation}
\red{A detailed derivation of Eq.~\eqref{Qpauli} from Eq.~\eqref{deltad2pauli} is presented in Appendix~\ref{appB}.}

The maximum value of decrease in the MSTD is \red{thus} given by
\begin{equation}
\Delta\overline{D^2} = \frac{2}{5}\max (\lambda_{\max} , 0)\,,
\end{equation}
where $\lambda_{\max}$ is the largest eigenvalue of  matrix $Q$ in Eq. (\ref{Qpauli}). The normalized eigenstate {$(x_0,~\mathbf{x})^T=\mathbf{e}_i$ with $\mathbf{e}_0 = (1,0,0,0)$, $\mathbf{e}_1 = (0,1,0,0)$, $\mathbf{e}_2 = (0,0,1,0)$, and $\mathbf{e}_3 = (0,0,0,1)$} corresponding to the largest eigenvalue will determine the quasi inverse, {as explained in the appendix}, which is given by
\begin{equation}
V  =  \sigma_i \quad  \text{ if ~} p_i = p_{\max}\,,\quad i\in \{1,2,3\}\,,\label{qi1}
\end{equation}
after using Eq. \eqref{defV}.
In this case, the decrease in the MSTD is given by
\begin{equation}
\Delta\overline{D^2} = \frac{2}{5}\max(p_{\max} - p_0 , 0)\,,\label{d1}
\end{equation}
where $p_{\max}=\max(p_1,p_2,p_3)$. Therefore, the quasi inverse of the Pauli channel exists \textit{if} $p_{\max}>p_0$. We find that if $p_0 \leq 1/2$ and $p_{\max}\geq 1/2$, then $\Delta\overline{D^2}$ is maximized.

\subsection{Generalized Amplitude Damping Channel}
The generalized amplitude damping channel $\mathcal{E}_{\text{GAD}}$ is a non-unital channel characterized by Krauss operators \cite{nielsen2010}
\begin{equation}
	\begin{aligned}
		E_1 &= \sqrt{p}\begin{pmatrix}
			1 & 0 \\
			0 & \gamma
		\end{pmatrix}\,, \\ 
		E_2 & = \sqrt{p}\begin{pmatrix}
			0 & \sqrt{1-\gamma^2} \\
			0 & 0
		\end{pmatrix}\,, \\
		E_3 &= \sqrt{1-p}\begin{pmatrix}
			\gamma & 0 \\
			0 & 1
		\end{pmatrix}\,, \\
		E_4 & = \sqrt{1-p}\begin{pmatrix}
			0 & 0 \\
			\sqrt{1-\gamma^2} & 0
		\end{pmatrix}\,.
	\end{aligned}
	\end{equation}

The affine map for this channel can be worked out as
\begin{equation}\label{gad_M}
	M = \begin{pmatrix}
		\gamma & 0 & 0 \\
		0 & \gamma & 0 \\
		0 & 0 & \gamma^2
	\end{pmatrix}  , \quad \mathbf{c} = \begin{pmatrix}
	0 \\
	0 \\
	(1-\gamma^2)(2p-1)
	\end{pmatrix}\,.
\end{equation}
\red{Note that the $p$-dependence is present only in the translational vector $\mathbf c$ and matrix $M$ is independent of $p$.}

\red{Using Eqs. \eqref{ms1} and \eqref{ms2} in Eq. \eqref{max}, we get
\begin{equation}\small
\begin{aligned}
\Delta \overline{D^2} &= \frac{1}{20}\left[\text{Tr}(MM^\dagger - NN^\dagger) -2\text{Tr}(M-N)\right] + \frac{1}{4}\left(\abs{\mathbf{c}}^2 - \abs{\mathbf u}^2\right) \\
& = \frac{1}{20}\left[\text{Tr}(MM^\dagger - NN^\dagger) -2\text{Tr}(M-N)\right] + \frac{1}{4} \left(\abs{\mathbf{c}}^2 - \abs{M^i \mathbf{c}}^2\right)
\end{aligned}
\end{equation}
Since the quasi inverse is a unitary map, $M^i \mathbf{c}$ is just a rotated version of $\mathbf{c}$, preserving the norm such that $\abs{\mathbf{c}}^2 = \abs{M^i \mathbf{c}}^2$, and thus
\begin{equation}
	\Delta\overline{ D^2} = \frac{1}{20}\left[\text{Tr}(MM^\dagger - NN^\dagger) -2\text{Tr}(M-N)\right]\,.
\end{equation}
Since  $M$ is indpendent of $p$, so is $N=M^iM$. Therefore, $\Delta\overline{ D^2} $ is independent of $p$.}

\red{
Using Eqs., \eqref{matrix_N}, \eqref{m_i}, and \eqref{gad_M}  in the above expression gives:
}
\begin{equation}
	\begin{gathered}
		\Delta \overline{D^2} = \frac{1}{10} \left[
		x_0^2 (\gamma^2 + 2\gamma) - 
		x_1^2\gamma^2 - 
		x_2^2\gamma^2 + \right. \\
		\left. x_3^2(\gamma^2 - 2\gamma) - 
		\gamma^2 - 2\gamma
		\right]
	\end{gathered}\,.
\end{equation}

{Using the identity $x_0^2 + \mathbf x\cdot \mathbf x = 1$}, \red{the above expression also takes the form of Eq. (\ref{d2matrix}) with}
\begin{equation}
Q = \frac{1}{2}\text{diag}\big(0,-\gamma(\gamma+1),-\gamma(\gamma+1),-2\gamma\big)\,.
\end{equation}
When $\gamma\geq0$, $\lambda_{\max}=0$ with corresponding eigenstate $\mathbf e_0  = (1,0,0,0)$ that results in $V=I$, that is, no quasi inverse exists for positive values of $\gamma$. Clearly, the largest eigenvalue $\lambda_{\max}=-{\gamma}$ for ${\gamma}<0$ with corresponding eigenstate $\mathbf e_3 = (0,0,0,1)$ that leads to $V=\sigma_3$ and thus 
\begin{equation}\label{qi2}
V= \begin{cases}
I\,, & \quad \text{for }\gamma \geq 0\,, \\
\sigma_3, & \quad \text{for }\gamma < 0\,,
\end{cases}
\end{equation}
in agreement with Ref. \cite{karimipour2020qubit}. 
In this case, the maximum  decrease in the MSTD is given by
\begin{equation}
\Delta\overline{D^2} = \frac{2}{5} \begin{cases}
0\,, & \quad \text{for }\gamma \geq 0\,, \\
-\gamma\,, & \quad \text{for }\gamma < 0\,.
\end{cases}\label{d2}
\end{equation}
For example, for $p=1$ and $\gamma=0$, $Q = \mathrm{diag}(0,0,0,0)$, which means $\Delta \overline{D^2} = 0$ for which $V=I$.

The generalized amplitude damping channel is composed of dissipation part and  the phase-flip part. The quasi inverse, \red{being a unitary map}, only reverses the phase-flip part by applying $\sigma_3$ when $\gamma$ is negative. So, the dissipation part of the channel cannot be reversed.

\subsection{Mixed Unitary Channel}
A mixed unitary channel is given by \cite{karimipour2020qubit}
\begin{equation}
	\mathcal{E} (\rho) = (1-3p)\rho + p\sum_{i}^{3} U_i \rho U_i^\dagger\,,
\end{equation}
where $U_i = \exp(-i\frac{\theta}{2}\sigma_i) = \cos(\frac{\theta}{2})I - \sin(\frac{\theta}{2})\sigma_i$ is a unitary operator that rotates the qubit around $x_i$ axis by an angle $\theta$. 

{
The affine map for this channel is found as
\begin{equation}\label{mixed_M}
	M = \begin{pmatrix}
		-q & - v & v \\
		v & - q & - v \\
		- v & v & -q
		\end{pmatrix} \quad , \quad \mathbf{c} = \begin{pmatrix}
		0 \\
		0 \\
		0
		\end{pmatrix}\,,
\end{equation}
where $v = p\sin(\theta)$ and $q = 4p\sin[2](\frac{\theta}{2})-1$. Using Eqs. \eqref{ms1}, \eqref{ms2}, \eqref{m_i}, and \eqref{mixed_M} in Eq.  \eqref{max} we get
\begin{equation}
	\begin{gathered}
		\Delta \overline{D^2} = \frac{2}{5} \left[
		  \frac{q}{4}\left(-3x_0^2 + x_1^2 + x_2^2 + x_3^3\right) + \right.\\
		\left.v(x_0x_1 + x_0x_2 + x_0x_3) + \frac{3q}{4}
		\right]\,.
	\end{gathered}
\end{equation}
}

\red{This allows us to write the above expression in the matrix form given by Eq. (\ref{d2matrix}) with}
\begin{equation}
	Q = \begin{pmatrix}
		0 & v/2 & v/2 & v/2 \\
		v/2 & q & 0 & 0 \\
		v/2 & 0 & q & 0 \\
		v/2 & 0 & 0 & q
	\end{pmatrix}\,.
\end{equation}

For $q \geqslant 0$, the largest eigenvalue of this matrix is $\lambda_{\max }=\frac{1}{2}(q+$ $\sqrt{q^2+3 v^2}$ ), with the corresponding eigenstate given by $\left(\frac{3 v}{2 \lambda_{\max }} \quad 1 \quad 1 \quad 1\right)^T$. This means that the quasi inverse of the channel is 
\begin{equation}
	V=e^{i \phi \mathbf{n} \cdot \mathbf{\sigma}}\,,\label{qi3}
\end{equation}
where 
\begin{equation}
\cos \phi=\frac{\sqrt{3} v}{\sqrt{3 v^2+4 \lambda_{\max }^2}}, \quad \mathbf{n}=\frac{1}{\sqrt{3}}(\hat{x}+\hat{y}+\hat{z})\,.
\end{equation}
The decrease in the MSTD is given by 
\begin{equation}
\Delta \overline{D^2}=\frac{2}{5} \lambda_{\max}\,.\label{d3}
\end{equation}

\subsection{Tetrahedron Channel}
This is the channel whose quasi inverse is different from one of its own Kraus operators. Its mathematical form is given as
\begin{equation}
	\mathcal{E}(\rho) = q\rho + \sum_{i=0}^3 p_i (\mathbf{v}_i\cdot \mathbf{\sigma}) \rho (\mathbf{v}_i\cdot \mathbf{\sigma})\,,
\end{equation}
where $$q = 1-\sum_{i=0}^3 p_i$$ and the vectors $\mathbf{v}_i$ are the corners of tetrahedron:
\begin{equation}
	\begin{aligned}
		\mathbf{v}_0 & = \frac{1}{\sqrt{3}} \begin{pmatrix}
			1 & 1 & 1
		\end{pmatrix} \,,\\
		\mathbf{v}_1 & = \frac{1}{\sqrt{3}} \begin{pmatrix}
			1 & -1 & -1
		\end{pmatrix}\,, \\
		\mathbf{v}_2 & = \frac{1}{\sqrt{3}} \begin{pmatrix}
			-1 & 1 & -1
		\end{pmatrix}\,, \\
		\mathbf{v}_3 & = \frac{1}{\sqrt{3}} \begin{pmatrix}
			-1 & -1 & 1
		\end{pmatrix}\,.
	\end{aligned}
\end{equation}

We consider a special case \cite{karimipour2020qubit} with
\begin{equation}
	p_1 = p_2 = p\,, \quad p_0 = p_3 = p^\prime \,,
\end{equation}
where $p+p^\prime\leq 0.5$ due to normalization of probability.

The affine map for this channel is found as
\begin{equation}\label{tetra_M}
\begin{gathered}
	M = \begin{pmatrix}
		- \frac{8 p}{3} - \frac{8 p^\prime}{3} + 1 & 
		- \frac{4 p}{3} + \frac{4 p^\prime}{3} & 
		0 \\
		- \frac{4 p}{3} + \frac{4 p^\prime}{3} & 
		- \frac{8 p}{3} - \frac{8 p^\prime}{3} + 1 
		& 
		0 \\
		0 & 
		0 & 
		- \frac{8 p}{3} - \frac{8 p^\prime}{3} + 1
	\end{pmatrix} \,,\\
\mathbf{c} = \begin{pmatrix}
				0 & 0 & 0
			\end{pmatrix}^T\,.
\end{gathered}
\end{equation}
Using Eqs. \eqref{ms1}, \eqref{ms2}, \eqref{m_i},  and \eqref{tetra_M} in Eq. \eqref{max} we get
 \begin{equation}
 	\Delta \overline{D^2} = \frac{2}{5} \left[ \left(\frac{8p}{3} + \frac{8p^\prime}{3} - 1\right)(x_1^2 + x_2^2 + x_3^2) + \frac{4}{3}(p^\prime - p)x_1x_2 \right]\,,
 \end{equation}
\red{which can be written in the matrix form given in Eq. (\ref{d2matrix}) with}
\begin{equation}
	Q = \begin{pmatrix}
		0 & 0 & 0 & 0 \\
		0 & \frac{8p}{3} + \frac{8p^\prime}{3} - 1 & -\frac{2p}{3}+ \frac{2p^\prime}{3} & 0 \\
		0 & -\frac{2p}{3}+ \frac{2p^\prime}{3} & \frac{8p}{3} + \frac{8p^\prime}{3} - 1 & 0 \\
		0 & 0 & 0 & \frac{8p}{3} + \frac{8p^\prime}{3} - 1
	\end{pmatrix}\,.
\end{equation}

The maximum value of decrease in MSTD is given by
\begin{equation}
	\Delta\overline{D^2} = \frac{2}{5}\begin{cases}
		\displaystyle{\max}\left\{2 p^\prime-1+\frac{10 p}{3}, 0\right\}\,, & \text { if } p \geq p^\prime\,, \\
		\displaystyle{\max}\left\{2 p-1+\frac{10 p^\prime}{3}, 0\right\}\,, & \text { if } p \leq p^\prime\,.
	\end{cases}\label{d4}
\end{equation}
with corresponding eigenstates given as
\begin{equation}
	(x_0,\mathbf x) = \begin{cases}
		\frac{1}{\sqrt{2}}(0,1,1,0) & \text { if } p \geq p^\prime\,, \\
		\frac{1}{\sqrt{2}}(0,1,-1,0) & \text { if } p \leq p^\prime\,.
	\end{cases}\label{x4}
\end{equation}
Therefore, the quasi inverse is
\begin{equation}
	V = \begin{cases}
		\displaystyle\frac{\sigma_1 + \sigma_2}{\sqrt{2}}\,, & \text{ if }\lambda_{\max}=2p^\prime -1 + \displaystyle\frac{10p}{3}\,, \\[5mm]
		\displaystyle\frac{\sigma_1 - \sigma_2}{\sqrt{2}}\,, & \text{ if }\lambda_{\max}=2p -1 + \displaystyle\frac{10p^\prime}{3}\,.
	\end{cases}\label{qi4}
\end{equation}

Let us note that the quasi inverses of the four example channels in Eqs. (\ref{qi1}), (\ref{qi2}), (\ref{qi3}), and (\ref{qi4}) have turned out to be the same as in Ref. \cite{karimipour2020qubit} using the fidelity. However, the expressions in Eqs. (\ref{d1}), (\ref{d2}), (\ref{d3}), and (\ref{d4}) for the decrease in the MSTD are different than those for pure states in Ref. \cite{karimipour2020qubit} but reduce to the same value if we had performed averaging just over the surface of the Bloch sphere.

\section{Conclusions}\label{sec5}
We proposed an alternative definition for quasi inverse of a channel that can easily extend the  mixed input states.  The quasi inverse based on maximizing the mean squared trace distance (MSTD) over the full Bloch ball of input state of a completely positive and trace preserving channel was postulated to be a unitary operator. This resulted in a constrained optimization program to find the four real parameters of the inverse operator. The results for Pauli,  generalized amplitude damping, mixed unitary, and the tetrahedron channels agreed with those derived based on the maximization of the average fidelity of a channel. {As an additional test for the proposed method, the quasi inverse of a general unitary operator was found to be equal to the  inverse of the unitary operator.} Therefore, this new definition provides an equivalent methodology to find the quasi inverse that is also generalizable to mixed states.

\appendix
\renewcommand{\theequation}{A.\arabic{equation}}
\setcounter{equation}{0}

\section{Maximization}\label{appA}
To maximize $f(\mathbf x) = \mathbf x^T Q \mathbf x$ subject to the constraint $\|\mathbf x \|^2=1$, we can use a Lagrange function 
\begin{equation}
	\begin{aligned}
		\mathcal{L}(\mathbf x , \lambda) &= \mathbf x^T Q \mathbf x - \lambda \left(\| \mathbf x \|^2 - 1\right)\,.
	\end{aligned}
\end{equation}
Next, we need to compute partial derivatives of $\mathcal{L}$ with respect to each component of $\mathbf x$ and $\lambda$ and set them equal to $0$ to get
\begin{equation}
	\begin{aligned}
		\frac{\partial \mathcal{L}}{\partial \mathbf x} & = 2Q\mathbf x - \lambda(2\mathbf x) = 0 \quad \Rightarrow \quad Q\mathbf x = \lambda \mathbf x\,, \\
		\frac{\partial \mathcal{L}}{\partial \lambda} & = \|\mathbf x \|^2 - 1 = 0 \quad \Rightarrow \quad \|\mathbf x \|^2 = 1\,.
	\end{aligned}
\end{equation}
Therefore, the problem of optimization of $f(\mathbf x) = \mathbf x^T Q \mathbf x$ has been converted to
an eigenvalue problem of matrix $Q$ and the solution of the optimization problem $\mathbf x$ and $\lambda$ must be an eigen pair of $Q$. Since $Q$ has four eigenstates $\mathbf x=\mathbf v_i$ with corresponding eigen values $\lambda = \lambda_i$, $i=1,2,3,4$. For each solution,  $f(\mathbf x)$ becomes
\begin{equation}
	f(\mathbf x) =  \mathbf v_i^T \left(\lambda_i \mathbf v_i\right) = \lambda_i \mathbf v_i^T\mathbf v_i = \lambda_i\,,\quad i=1,2,3,4.
\end{equation}
Thus
\begin{equation}
	\max f(\mathbf x) = \max \left(\lambda_1, \lambda_2, \lambda_3,\lambda_4\right)\,,
\end{equation}
and the eigenstate $\mathbf v_j$ corresponding to $\lambda_j=\lambda_{\max}$ is the global maximizer of $f(\mathbf x)$.

\section{A Sample Derivation}\label{appB}
\red{
We now derive Eq.~\eqref{Qpauli} from Eq.~\eqref{deltad2pauli}.
}
\red{
Gathering the coefficients of $x_i^2$ in Eq.~\eqref{deltad2pauli} and using the  identity $\sum_{i=0}^3 p_i = 1$, we get:
\begin{equation*}
	\Delta\overline{D^2} = \frac{2}{5} \left[\sum_{i=0}^3 \left(p_i - \frac{1}{4}\right)x_i^2 + \left(\frac{1}{4}-p_0\right) \right] \,,
\end{equation*}
which can be rearranged in matrix form as
\begin{equation*}
	\begin{aligned}
		\Delta \overline{D^2} &= \frac{2}{5} \begin{pmatrix}
			x_0 & \mathbf{x}^T
		\end{pmatrix} 
		\begin{pmatrix}
			p_0 - \frac{1}{4} & 0 & 0 & 0 \\
			0 & p_1 - \frac{1}{4} & 0 & 0 \\
			0 & 0 & p_2 - \frac{1}{4} & 0 \\
			0 & 0 & 0 & p_3 - \frac{1}{4}
		\end{pmatrix}\begin{pmatrix}
			x_0 \\
			\mathbf{x}
		\end{pmatrix} \\ 
		& + \frac{2}{5}\left(\frac{1}{4} - p_0 \right)\begin{pmatrix}
			x_0 & \mathbf{x}^T
		\end{pmatrix}
		\mathbb{I}_{4\times 4}
		\begin{pmatrix}
			x_0 \\
			\mathbf{x}
		\end{pmatrix} \\
		& = \frac{2}{5} 
		\begin{pmatrix}
			x_0 & \mathbf{x}^T
		\end{pmatrix} Q
		\begin{pmatrix}
			x_0 \\
			\mathbf{x}
		\end{pmatrix}
	\end{aligned}
\end{equation*}
where $Q=\mathrm{diag}(0,p_1-p_0,p_2-p_0,p_3-p_0)$ and $\mathbb I_{4\times 4}$ is $4\times 4$ identity matrix. A similar derivation holds for other examples as well.
}


\begin{thebibliography}{99}
\bibitem{nielsen2010}
M. A. Nielsen and I. L. Chuang,
Quantum Computation and Quantum Information,
Cambridge University Press, (2010).


\bibitem{karimipour2020qubit}
V. Karimipour, F. Benatti, and R. Floreanini,
Quasi-inversion of qubit channels,
Phys. Rev. A, 101, 032109, (2020).



\bibitem{shahbiegi2021finite}
F. Shahbeigi, K. Sadri, M. Moradi, K. Życzkowski, and V. Karimipour,
Quasi-inversion of quantum and classical channels in finite dimensions,
J. Phys. A.: Math. Theor., 54, 345301, (2021).

\bibitem{karimipour2022on}
V. Karimipour,
On quasi-inversion of quantum channels in 2 and in higher dimensions,
Open Sys. Inf. Dynamics, 29, 2250014, (2022).


\bibitem{cao2022neural}
N. Cao, J. Xie, A. Zhang, S.-Y. Hous, L. Zhang, and B. Zeng,
Neural networks for quantum inverse problems,
New J. Phys. 24, 063002, (2022).

\bibitem{vesperini2023ent}
A. Vesperini, G. Bel-Hadj-Aissa, and R. Franzosi,
Entanglement and quantum correlation measures for quantum multipartite mixed states,
Sc. Rep. 13, 2852, (2023).

\bibitem{ghosh2018quantum}
S. Ghosh and S. Raju,
Quantum information measures for restricted sets of observables,
Phys. Rev. D, 98, 046005, (2018).

\bibitem{wilde2013}
M. M. Wilde,
Quantum Information Theory,
Cambridge University Press, (2013).





\end{thebibliography}
\end{document}